\documentclass[aps,pra,twocolumn,superscriptaddress,floatfix,
nofootinbib,showpacs,longbibliography]{revtex4-1}
\usepackage[utf8]{inputenc}  
\usepackage[T1]{fontenc}     
\usepackage[british]{babel}  
\usepackage[sc,osf]{mathpazo}
\usepackage[scaled=0.86]{berasans}  
\usepackage[colorlinks=true, citecolor=blue, urlcolor=blue]{hyperref}  
\usepackage{graphicx}
\usepackage[babel]{microtype}  
\usepackage{amsmath,amssymb,amsthm,bm,amsfonts,mathrsfs,bbm} 

\usepackage{xspace}  
\usepackage{pgf,tikz}
\usepackage{xcolor}
\usepackage{multirow}
\usepackage{array}
\usepackage{bigstrut}
\usepackage{braket}
\usepackage{color}
\usepackage{natbib}
\usepackage{multirow}
\usepackage{mathtools}
\usepackage{float}
\usepackage{xcolor,colortbl}
\usepackage{physics}
\usepackage{amsmath}
\usepackage{color}
\usepackage[justification=justified, format=plain]{subcaption}
\usepackage[justification=raggedright]{caption}

\newcommand{\be}{\begin{equation}}
\newcommand{\ee}{\end{equation}}
\newcommand{\ba}{\begin{eqnarray}}
\newcommand{\ea}{\end{eqnarray}}

\def\>{\rangle}
\def\<{\langle}

\begin{document}
	
\title{Higher-dimensional entanglement detection and  quantum channel characterization using moments of generalized positive maps}

\author{Bivas Mallick}
\email{bivasqic@gmail.com}
\affiliation{S. N. Bose National Centre for Basic Sciences, Block JD, Sector III, Salt Lake, Kolkata 700 106, India}

\author{Ananda G. Maity}
\email{anandamaity289@gmail.com}
\affiliation{Networked Quantum Devices Unit, Okinawa Institute of Science and Technology Graduate University, Onna-son, Okinawa 904 0495, Japan}

\author{Nirman Ganguly}
\email{nirmanganguly@hyderabad.bits-pilani.ac.in}
\affiliation{Department of Mathematics, Birla Institute of Technology and Science Pilani, Hyderabad Campus, Hyderabad, Telangana-500078, India}

\author{A. S. Majumdar}
\email{archan@bose.res.in}
\affiliation{S. N. Bose National Centre for Basic Sciences, Block JD, Sector III, Salt Lake, Kolkata 700 106, India}

\begin{abstract}
Higher-dimensional entanglement is a valuable resource for several quantum information processing tasks, and is often characterized by the Schmidt number and specific classes of entangled states beyond qubit-qubit and qubit-qutrit systems. We propose a criterion to detect higher-dimensional entanglement, focusing on determining the Schmidt number of quantum states and identifying
significant classes of PPT and NPT entangled states. Our approach relies on evaluating moments of generalized positive maps which can be efficiently simulated in real experiments without the requirement of full-state tomography. We demonstrate the effectiveness of our detection scheme through various illustrative examples. As a direct application, we explore the implications of our moment-based detection schemes in identifying useful quantum channels such as non-Schmidt number breaking channels and non-entanglement breaking channels. Finally, we present an operational implication of our proposed moment criterion through its manifestation in channel discrimination tasks.
\end{abstract}
\maketitle

\section{Introduction}
Quantum entanglement plays a pivotal role in the rapidly growing field of quantum technologies \cite{peres1997quantum,einstein1935can,horodecki2009quantum,plenio2014introduction,nielsen2010quantum}. It underpins numerous quantum information processing tasks such as superdense coding \cite{bennett1992communication}, quantum teleportation \cite{bennett1993teleporting}, quantum cryptography \cite{ekert1991quantum,branciard2012one}, quantum secret sharing \cite{hillery1999quantum,cleve1999share,bandyopadhyay2000teleportation}, remote state preparation \cite{pati2000minimum}, and many others \cite{horodecki2009quantum}. Yet, before harnessing entanglement for quantum information processing tasks, it is essential to verify whether entanglement has indeed been established between the parties. A conventional method for entanglement detection is the positive partial transposition (PPT) criterion which provides a necessary and sufficient condition for the detection of entanglement in qubit-qubit and qubit-qutrit scenarios \cite{horodecki2001separability}. However, this criterion has limitations, as PPT entangled states can exist in higher-dimensional systems, which remain undetected by this approach. Though, identifying whether an unknown quantum state is entangled or not has been proven to be NP-hard \cite{gurvits2004classical,gharibian2010strong,gharibian2015quantum}, developing reliable methods to detect entanglement in arbitrary-dimensional quantum systems, including PPT entangled states remains an active area of research \cite{lewenstein2001characterization,spengler2012entanglement, shang2018enhanced,sarbicki2020family,terhal2000bell,guhne2009entanglement, ganguly2009witness,ganguly2013common,ganguly2014witness,goswami2019universal, bhattacharya2021generating,mallick2024genuine}.

Related to the entanglement detection problem is the identification of the minimal dimension needed to reproduce the correlations within quantum states. This leads to the concept of the entanglement dimension, which is characterized by the Schmidt number (SN) of a bipartite density matrix \cite{terhal2000schmidt}. The Schmidt number actually quantifies the number of levels that contribute to the generation of entanglement between the particles.
Recent studies have demonstrated that states with higher Schmidt numbers offer significant advantages in several quantum information processing tasks, such as channel discrimination \cite{bae2019more}, power of quantum communication \cite{zhang2025quantum}, quantum control \cite{kues2017chip}, and security of quantum key distribution \cite{cerf2002security}. Despite numerous applications of Schmidt number, a fundamental challenge is determining the Schmidt number of a quantum state to effectively use it as a resource in quantum information processing tasks. 

Several methods have been proposed in the literature to detect  Schmidt number, based on different perspectives and properties of a quantum state \cite{terhal2000schmidt,liu2023characterizing,liu2024bounding, tavakoli2024enhanced,sanpera2001schmidt,wang2024schmidt,shi2024families, wyderka2023construction,bavaresco2018measurements,liang2025detecting, krebs2024high,yang2016all,zhang2024analyzing,mukherjee2025measurement,engineer2024certifying}.
In this work, one of our main focuses is to detect Schmidt number even when partial knowledge about the state is available. The concept of partial transpose moments (PT moments) to characterize correlations in many-body systems was
introduced earlier \cite{calabrese2012entanglement}. Building on this idea, 
a criterion for entanglement detection using PT moments was developed \cite{elben2020mixed}, offering a practical approach that overcomes experimental challenges in entanglement detection. Inspired by these advancements, in this work, we aim to propose a resource efficient detection scheme for identifying the signature of higher-dimensional entanglement through the moments of generalized positive maps.

We begin by defining the moments of a generalized positive map and then explore how these moments can be exploited to efficiently detect higher-dimensional entanglement-- specifically, the Schmidt number of quantum states and specific classes of PPT and NPT (negative partial transpose) entangled states in qutrit-qutrit systems. While certain classes of PPT entangled states have previously been detected using moments of the Choi map \cite{wang2022operational}, in this work we demonstrate that the moments of the Breuer–Hall map enable the detection of specific classes of PPT entangled states that are not identifiable using moments of the Choi map. Our method, based on these moment criteria, involves evaluating simple functionals that can be efficiently calculated in real experiments through a technique called shadow tomography \cite{aaronson2018shadow,aaronson2019gentle,huang2020predicting,elben2020mixed}. This approach is more resource-efficient than the usual full-state tomography. Furthermore, our moment-based approach is more accessible in terms of the number of state copies needed to estimate them. As system size increases, while the usual full tomography demands an exponentially growing number of measurements, these moment-based methods only require a polynomial number of state copies. Moreover, our criterion does not rely on any prior knowledge of the state, unlike witness-based detection methods, which are state-dependent.

Next, we explore the significance and implications of our proposed detection schemes in the context of quantum channels. In some cases, environmental noise can be so intense that the quantum state loses its value as a resource \cite{srinidhi2024quantum,muhuri2023information,heinosaari2015incompatibility,horodecki2003entanglement}. There are certain classes of quantum channels that completely destroy the entanglement between the subsystem they act upon from the rest of the system. That is, regardless of the initial state, the application of these channels always produces a separable state. Such channels are known as entanglement breaking channels \cite{horodecki2003entanglement}. There exists a broader class of channels that generalizes the concept of entanglement-breaking channels. These channels are characterized by their ability to reduce the Schmidt number of a bipartite composite state which makes them resource-breaking. Such channels are referred to as Schmidt number breaking channels \cite{chruscinski2006partially,devendra2023mapping,mallick2024characterization}. As a direct
application of our state detection schemes, here we
aim to detect the signature of such quantum channels
that are non-entanglement-breaking and non-Schmidt number breaking, as these channels are crucial for preserving and utilizing entanglement resources in various quantum information processing tasks, particularly in quantum communication and distributed computing \cite{zhang2025quantum,cozzolino2019high}. Further, we also demonstrate an operational implication of our proposed moment criteria through its manifestation in channel discrimination tasks.

The rest of the paper is organized as follows. In section \ref{s2}, we provide a brief overview of the essential preliminaries concerning generalized positive maps, Schmidt number, entanglement breaking and Schmidt number breaking channels, and minimum error discrimination as well as the moment criteria proposed in earlier works for entanglement detection. In section \ref{s3}, we present our framework for detection of several fundamental quantities in quantum information theory, such as the Schmidt number, PPT entanglement, NPT entanglement. Section \ref{s4} explores the implications of our moment-based detection schemes, particularly in identifying useful quantum channels (e.g., non-Schmidt number breaking channels) and performing quantum channel discrimination tasks. Finally, in section \ref{s5}, we summarize our main findings and outline possible future directions.

\section{Preliminaries}\label{s2}
Here, we present the necessary background and mathematical primitives that are essential to follow the rest of the work.

\subsection{Structure of positive maps in $\mathcal{M}_d$}
Consider a bipartite system composed of two subsystems, $A$ and $B$ each associated with a $ d$-dimensional complex Hilbert space ${\mathbf{C}}^d$. The composite system is described by the tensor product Hilbert space ${\mathbf{C}}^d \otimes {\mathbf{C}}^d$.
Quantum states in this Hilbert space are represented by density operators, which are positive operators with unit trace.  The set of all such density operators is denoted by $\mathcal{D} ({\mathbf{C}}^d \otimes {\mathbf{C}}^d)$. Note that the operators acting on a finite-dimensional Hilbert space are bounded and can be expressed as matrices with respect to some suitable basis. Let, $\mathcal{M}_d$ and $\mathcal{M}_k$ denote the sets of complex matrices of dimensions $d \times d$ and $k \times k$ respectively. A linear map $\Lambda: \mathcal{M}_d \rightarrow \mathcal{M}_d $ is said to be positive, if $\Lambda(\rho) \ge 0$, for all $\rho \in \mathcal{M}_d$. A linear map $\Lambda: \mathcal{M}_d \rightarrow \mathcal{M}_d$ is said to be $k$-positive if the extended map $id_A \otimes \Lambda :\mathcal{M}_k \otimes \mathcal{M}_d \rightarrow \mathcal{M}_k \otimes \mathcal{M}_d$ is positive for some $k \in \mathbf{N}$. A linear map $\Lambda: \mathcal{M}_d \rightarrow \mathcal{M}_d $ is said to be completely positive if $id_A \otimes \Lambda : \mathcal{M}_k \otimes \mathcal{M}_d \rightarrow \mathcal{M}_k \otimes \mathcal{M}_d$ is positive for all $k \in \mathbf{N}$. To determine whether a positive map $\Lambda$ is completely positive, one can rely on the Choi-Jamiołkowski isomorphism \cite{jamiolkowski1974effective, choi1975positive}. The Choi matrix associated with the positive map $\Lambda$ is defined as $ {\mathcal{C}}_{\Lambda}= (id_A \otimes \Lambda) (\ket{{\phi}^+}\bra{{\phi}^+})$, with $\ket{{\phi}^+}=\frac{1}{\sqrt d} \sum_{i} \ket{ii}$ being the maximally entangled state in ${\mathbf{C}}^d\otimes {\mathbf{C}}^d$. A positive map $\Lambda$ is said to be completely positive if and only if the corresponding Choi matrix ${\mathcal{C}}_{\Lambda}$ is positive semidefinite. Moreover, a linear map $\Lambda$ is trace-preserving if $\text{Tr} (\Lambda (\rho)) = \text{Tr}(\rho)$ for all $\rho \in  \mathcal{M}_d$. A linear map $\Lambda$ is said to be trace-annihilating if $\text{Tr} (\Lambda (\rho)) = 0$ for all $\rho \in  \mathcal{M}_d$. \\

 Now, any positive map $\Lambda_{\mathcal{x}}: \mathcal{M}_d \rightarrow \mathcal{M}_d$ can always be decomposed as \cite{augusiak2009positive}:
\begin{equation} \label{generalform}
     {\Lambda}_{\mathcal{x}} (\rho) = \mu \text{Tr}(\rho) I_d - \Phi \hspace{.2cm} \text{for} \hspace{.2cm} \rho \in {\mathcal{M}}_d 
\end{equation}
where $\mu = d \lambda_{max}$ with $\lambda_{max}$ being the largest eigenvalue of the Choi matrix ($(id_A \otimes \Lambda_{\mathcal{x}}) (\ket{{\phi}^+}\bra{{\phi}^+})$) and $\Phi$ is some completely positive map. Several positive maps, including the Choi map, Breuer-Hall map, and Reduction map, which we are going to employ for detecting higher-dimensional entanglement can be expressed in the above form with an appropriate choice of $\Phi$. 
\begin{itemize}
\item \textbf{Reduction map:} The Reduction map is defined by its action as follows \cite{terhal2000schmidt,augusiak2009positive}:
\begin{equation} \label{reduction}
    {\Lambda}_{R} ( \rho) := \text{Tr}( \rho) I_d - k  \rho \hspace{.2cm} \text{for} \hspace{.2cm}  \rho \in {\mathcal{M}}_d. 
\end{equation}
This map ${\Lambda}_R ( \rho)$ is $r$-positive but not $r+1$-positive (where, $r <d $) for 
\begin{equation} \label{rpositive}
    \frac{1}{r+1} < k \le \frac{1}{r} .
\end{equation} 

\item \textbf{Breuer-Hall map:} The action of the Breuer-Hall map is defined as \cite{augusiak2009positive}: 
\begin{equation}\label{BH}
    \Lambda_{BH} ( \rho) := \text{Tr}(\rho) I_d - \rho- U \rho^{T} U^{\dagger} 
\end{equation}
for $ \rho \in {\mathcal{M}}_d$, and $U$ is an anti-symmetric matrix {\it i.e.} $U^{\dagger} = -U$, satisfying $UU^{\dagger} \le \mathbf{I}$.

\item \textbf{Generalized Choi map:} The generalized Choi map is defined as \cite{augusiak2009positive}:
\begin{equation} \label{choi}
 \Lambda_{C}^{d,k} ( \rho) := (d-k) \epsilon ( \rho) + \sum_{i=1}^{k}  \epsilon (S^{i} \rho {S^{i}}^{\dagger}) - \rho
\end{equation}
for $ \rho \in {\mathcal{M}}_d$, where $ \epsilon$ is a completely positive map defined as,
\begin{equation}
    \epsilon ( \rho) = \sum_{i=0}^{d-1} \bra{j} \rho\ket{j} \ket{j} \bra{j}
\end{equation}
and $S^{i} \ket{i} = \ket{i-1}$ (mod $d$). For $k=d-1$, Eq.~\eqref{choi} reduces to the Reduction map defined in Eq.~\eqref{reduction}.
For $k=1,2,..,d-2$, above map defined in Eq.~\eqref{choi} has been shown to be indecomposable \cite{ha1998atomic}. In particular for $d=3$ and $k=1$, this map reduces to the well-known Choi map \cite{choi1975positive} which is defined as 
\begin{equation} \label{choimap}
  \Lambda_{C}^{3,1} ( \rho) := \begin{bmatrix} 
	\rho_{11} + \rho_{22}& -\rho_{12} &-\rho_{13}\\  
	-\rho_{21}& \rho_{22}+ \rho_{33}& -\rho_{23}\\
    -\rho_{31}& -\rho_{32}&  \rho_{33}+\rho_{11}\\
		\end{bmatrix}
\end{equation}
for all  \begin{equation}
  \rho= \begin{bmatrix} 
	\rho_{11} & \rho_{12} &\rho_{13}\\  
	\rho_{21}& \rho_{22}& \rho_{23}\\
    \rho_{31}& \rho_{32}&  \rho_{33}\\
		\end{bmatrix}\nonumber \in \mathcal{M}_3 \text{ and } \{\rho_{ij}\}\in \mathbb{C}.
\end{equation}
\end{itemize}

In Table~\ref{notationstable}, we demonstrate that for certain values of $\mu$ and $\Phi$, all these maps can be obtained from the general form given in Eq.~\eqref{generalform}.

\begin{table}[ht]
\centering
\begin{tabular}{| c| c| c| } 

\hline
 \textbf{Map}  &   \textbf{$\mu$}  &   \textbf{$\Phi$}\\
 \hline
Reduction map $( {\Lambda}_{R})$ & 1 & $ k\rho$ \\
\hline
Breuer-Hall map $( {\Lambda}_{BH})$& 2 & Tr$(\rho) I_d + \rho + U{\rho}^{T} U^{\dagger}$ \\
\hline
Generalized Choi map $(\Lambda_{C}^{d,k})$ & $d-k$ & $(d-k) \text{Tr} (\rho)$ $I_d$ -  $\Lambda_{d,k} ( \rho)$ \\
 \hline
\end{tabular}\\
\caption{The Reduction map, Breuer-Hall map, and Generalized Choi map are specific cases of the positive map defined in Eq.~\eqref{generalform}, obtained by specifying the value of $\mu$ and $\Phi$.}   
\label{notationstable}
\end{table}


\subsection{Schmidt number}
A bipartite pure state $\ket{\psi} \in {\mathbf{C}}^d \otimes {\mathbf{C}}^d$ can always be expressed in its Schmidt decomposition form \cite{peres1997quantum,nielsen2010quantum} as:
\begin{equation}
    \ket{\psi}= \sum_{i=1}^r \sqrt{{\lambda}_i} {\ket{i}}_A {\ket{i}}_B 
\end{equation}
where, ${\lambda}_i \ge 0$, $ \sum_i {\lambda}_i =1$, and $\ket{i_A} (\ket{i_B})$ forms an orthonormal basis in ${\mathcal{H}}_A ({\mathcal{H}}_B)$. Here, $r$ indicates the Schmidt rank of the pure state $ \ket{\psi}$. Later, Terhal and Horodecki \cite{terhal2000schmidt} extended this concept to mixed states by introducing the notion of Schmidt number. A bipartite density matrix $\rho$ has Schmidt number $r$, if in every possible decomposition of $\rho$ into pure states, {\it i.e.} \begin{equation}
    \rho= \sum_{k} p_k {\ket{{\psi}_k}} {\bra{{\psi}_k}}
\end{equation} with $p_k \ge 0$, at least one of the pure states ${\ket{{\psi}_k}}$ has Schmidt rank $r$. Moreover, there must exist at least one decomposition of $\rho$  in which all vectors  $\{\ket{{\psi}_k} \}$ have a Schmidt rank not greater than $r$. This can be expressed mathematically as:
\begin{equation}
    \text{SN} (\rho) := \min_{\rho = \sum_{k} p_{k}  \ket{{\psi}_k} \bra{{\psi}_k}}  \{\max_{k} \hspace{0.1cm} \text{SR}(\ket{{\psi}_k})\}
\end{equation}
where, the minimization is taken over all possible pure state decomposition of $\rho$ and SR $(\ket{{\psi}_k})$ represents the Schmidt rank of the pure state $\ket{{\psi}_k}$. For a bipartite state $\rho \in  \mathcal{D}({{\mathbf{C}}^d \otimes  {\mathbf{C}}^d})$, the Schmidt number satisfies $1 \le$ \hspace{0.08cm}SN$(\rho) \le d$. If 
$\rho$ is separable, then SN$(\rho) =1$. Let  $\mathcal{S}_r$ denote the set of all states in $\mathcal{D} ({\mathbf{C}}^d \otimes {\mathbf{C}}^d)$ whose Schmidt number is at most $r$. The set $\mathcal{S}_r$ forms a convex and compact subset within the space of density matrices $\mathcal{D} ({\mathbf{C}}^d \otimes {\mathbf{C}}^d)$. Furthermore, the sets satisfy the nested relation $\mathcal{S}_1  \subset \mathcal{S}_2 ...\subset \mathcal{S}_r$, where $\mathcal{S}_1$ 
  represents the set of separable states.

The set $\mathcal{S}_1$ is fully characterized by positive maps \cite{terhal2000schmidt,sanpera2001schmidt}. For $r >1$, characterizing $\mathcal{S}_r$ requires $r$-positive but not $r+1$-positive maps. Specifically, any such map $({\Lambda}_k)$ must satisfy the conditions: $(id_A \otimes {\Lambda}_k )(\rho) \ge 0 \hspace{0.2cm}$ for all $\rho \in \mathcal{S}_r$ and $(id_A \otimes {\Lambda}_k )(\sigma) < 0$ for at least one $ \sigma \notin \mathcal{S}_r$. One such example of a $r$-positive but not $r+1$-positive map is the Reduction map defined in Eq.~\eqref{reduction} for a specific range of parameter $k$. \\

\subsection{Entanglement breaking and Schmidt number breaking channels}
In the resource theory of entanglement, entangled states are considered as valuable resources. A particular class of quantum channels possesses the ability to completely disentangle the subsystem on which they act on from the rest of the system. Specifically, applying these channels to a subsystem of an arbitrary quantum state ensures that the resulting state is separable. Such channels are referred to as entanglement-breaking channels $(\mathbb{EB})$. Mathematically, a quantum channel $\mathcal{E}$ is said to be entanglement breaking, if $(id_A \otimes \mathcal{E}) (\rho)$ is separable for all $\rho$ \cite{horodecki2003entanglement}.  Furthermore, it was established in Ref.~\cite{horodecki2003entanglement} that a quantum channel $\mathcal{E}: \mathcal{M}_d \rightarrow \mathcal{M}_d $ is said to be entanglement breaking iff the corresponding Choi operator ${\mathcal{C}}_{\mathcal{E}}$ is separable \cite{horodecki2003entanglement}.\\

On the other hand, there are certain classes of quantum channels that are capable of reducing the Schmidt number of a bipartite composite state. These channels are known as Schmidt number breaking channels. Mathematically, A quantum channel $\mathcal{E}$ is classified as an $r$-Schmidt number breaking channel ($r-\mathbb{SNBC}$), if SN$[(id_A \otimes \mathcal{E}) \rho] \le r$, for all $\rho \in  \mathcal{D}({{\mathbf{C}}^d \otimes  {\mathbf{C}}^d})$, where $r < d$ \cite{chruscinski2006partially,mallick2024characterization}. Authors in \cite{chruscinski2006partially} proved that, a quantum channel $(\mathcal{E})$ belongs to $r-\mathbb{SNBC}$ iff SN$({\mathcal{C}}_{\mathcal{E}}) \le r$, where ${\mathcal{C}}_{\mathcal{E}}$ is the corresponding Choi state defined earlier. Note that every entanglement-breaking (EB) channel is  $1$-Schmidt number breaking channel ($1-\mathbb{SNBC}$). However, there exists $r-\mathbb{SNBC}$ with $r>1$ that are not $\mathbb{EB}$ \cite{mallick2024characterization}. This establishes that the set of $\mathbb{EB}$ channels is a strict subset of  $r-\mathbb{SNBC}$.

\subsection{Minimum error discrimination}\label{s2D}
Consider a preparation device that prepares a quantum system in one of many possible states $\rho_{k}$, where each preparation is associated with a probability $p_k$. The minimum-error quantum state
discrimination task aims at identifying the state correctly with the maximum probability of success (or equivalently, with minimum probability of error) through an optimal choice of measurements. The optimal probability of correctly guessing the state is given by, 
\begin{equation} \label{minimumerror}
    p_{success} (p_k,\rho_k)= \max_{M_k} \sum_k p_k \text{Tr}( M_k\rho_k) =  p_{success} (\tilde{\rho_k})
\end{equation}
where, $\tilde{\rho_k} = p_k \rho_k$ and the maximization is over all generalized measurements $M_k$. In general, $\{M_k\}$'s are positive-operator-valued measurement (POVM) such that $M_k \ge 0$ $\forall k$ and $\sum_k M_k = I_d $. The optimal discrimination strategy consists of choosing the optimal measurements that achieve this task. For two quantum
states $\rho_1$ and $\rho_2$ associated with probabilities $p$ and $1-p$ respectively, the optimal discrimination strategy with the corresponding optimal success probability has been obtained in \cite{helstrom1969quantum}. Taking $\tilde{\rho_1} = p_1 \rho_1$ and $\tilde{\rho_2} = p_2 \rho_2$, we have 
\begin{equation}
     p_{success} (\{ \tilde{\rho_1}, \tilde{\rho_2}\}) = \frac{1}{2} (1+ || \rho_1- \rho_2||_1 )
\end{equation}
where $||.||_1$ denotes the trace norm defined by $|| A||_1 = \text{Tr} \sqrt{{A}^{\dagger} A}. $

\subsection{Moment criteria}
Entanglement is a fundamental quantum phenomenon that serves as the foundation for numerous information processing applications. In the context of bipartite systems, a widely recognized method for detecting entanglement relies on the PPT (Positive Partial Transpose) criterion. This approach involves checking whether the partially transposed state $\rho^{T_A}_{AB}$ is positive semi-definite. If not, then the given state $\rho_{AB}$ is said to be entangled. This PPT criterion is both necessary and sufficient for detecting entanglement in ${\mathbf{C}}^2 \otimes {\mathbf{C}}^2$, ${\mathbf{C}}^2 \otimes {\mathbf{C}}^3$ and ${\mathbf{C}}^3 \otimes {\mathbf{C}}^2$ systems. However, calculating the full spectrum of eigenvalues for the partially transposed state $\rho^{T_A}_{AB}$ is not feasible in practical experiments because of its computational demands. To address this issue, Calabrese \textit{et al.} introduced the concept of moments of the partially transposed density matrix (PT-moments)  \cite{calabrese2012entanglement}.\\

 For a bipartite state $\rho_{AB}$, these $n$-th order partial transpose (PT)-moments \cite{gray2018machine,elben2020mixed,yu2021optimal,Neven2021,mallick2024assessing,mallick2025efficient,wang2022operational} are defined as follows:
\begin{equation}
p_n := \text{Tr}[{(\rho_{AB}}^{T_A})^n] \label{PTmoments}
\end{equation}
for n=1,2,3,.... Elben et. al. \cite{elben2020mixed}  proposed a simple yet effective criterion for detecting entanglement using only the first three moments. According to their criterion, if a state $\rho_{AB}$ is PPT, then ${p_2}^2 \leq p_3 p_1 $. Therefore, if a state  $\rho_{AB}$ violates this inequality, then the state is NPT and hence entangled. This criterion is commonly referred to as $p_3$-PPT criterion. For the detection of Werner state this $p_3$-PPT criterion is equivalent to the PPT criterion. Hence, it serves as a necessary and sufficient condition for detecting bipartite entanglement of Werner states. 
However, each higher order moment ($n \geq 4$) can give rise to an independent and different entanglement detection criterion \cite{yu2021optimal}. In Ref.~\cite{yu2021optimal}, the authors introduce the concept of Hankel matrices, denoted by $[H_{n}(\mathbf{p})]_{ij}$, where $i,j \in \{0, 1, ..., k\}$ and $\mathbf{p}=(p_1, p_2, ..., p_n)$. These matrices are defined as $(n+1) \times (n+1)$ matrices with elements defined by 
\begin{equation}
    [H_{n}(\mathbf{p})]_{ij} := p_{i+j+1}.  \label{Hankelmatrices}
\end{equation}
Hence, the first and the second Hankel matrices are expressed as
\begin{equation}
    H_1 = \begin{pmatrix}
p_1 & p_2   \vspace{0.2cm}\\ 
p_2 & p_3   \label{firstHankelmatrix}
\end{pmatrix} 
\end{equation}
    and 
\begin{equation}
    H_2 = \begin{pmatrix}
p_1 & p_2 & p_3  \vspace{0.2cm}\\ 
p_2 & p_3 & p_4  \vspace{0.2cm}\\
p_3 & p_4 & p_5      \label{secondHankelmatrix}
\end{pmatrix}
\end{equation} 
respectively. A necessary condition for separability based on Hankel matrices is given by 
    \begin{equation}
        \det[H_{n}(\mathbf{p})] \ge 0 . \label{Hankelmatrixcondition}
    \end{equation} 
These PT-moments can be experimentally obtained using shadow tomography, which bypasses the need for full state tomography, thereby significantly reducing resource consumption \cite{aaronson2018shadow,aaronson2019gentle,huang2020predicting,elben2020mixed,cieslinski2024analysing,liang2024real}.

Motivated by the above considerations, in the next section we explore how moment-based detection schemes can be developed for the detection of higher-dimensional entangled states, including Schmidt number of a quantum state and significant classes of PPT and NPT entangled states. First we define the moments of generalized positive maps $(s_n)$, and based on it we develop a formalism for detection of Schmidt number of a quantum state, as well as PPT and NPT entangled states.

\section{Detection of higher-dimensional entanglement} \label{s3}

\textbf{Definition 1:} Consider a linear, positive map $\Lambda_{\mathcal{x}}$ of the form given in Eq.~\eqref{generalform}. The $n$-th order moments, $s_n$ of the positive map, $\Lambda_{\mathcal{x}}$ are formally defined as:
\begin{equation} \label{moments}
s_n := \text{Tr}[S_{\mathcal{x}}^n] 
\end{equation}
where, 
\begin{equation}\label{def_S}
    S_{\mathcal{x}} = \frac{(id_A \otimes \Lambda_{\mathcal{x}})(\rho_{AB})}{\text{Tr}[(id_A \otimes \Lambda_{\mathcal{x}})(\rho_{AB})]}
\end{equation}  with $n$ being an integer.
\begin{itemize}
\item If we consider $\Lambda_{\mathcal{x}}$ to be $\Lambda_{R}$ for $\frac{1}{r+1} < k \le \frac{1}{r}$ (defined in Eq.~\eqref{reduction}),  we call Eq.~\eqref{moments} as the moments of $r$ positive but not $r+1$ positive Reduction map.

\item If we consider $\Lambda_{\mathcal{x}}$ to be $\Lambda_{R}$ for $ k =1$ (defined in Eq.~\eqref{reduction}),  we call Eq.~\eqref{moments} as the moments of Reduction map.

\item If we consider $\Lambda_{\mathcal{x}}$ to be $\Lambda_{BH}$ (defined in Eq.~\eqref{BH}), we call Eq.~\eqref{moments} as the moments of the Breuer–Hall map.

\item If we consider $\Lambda_{\mathcal{x}}$ to be $\Lambda^{3,1}_{C}$ (defined in Eq.~\eqref{choimap}), we call Eq.~\eqref{moments} as the moments of the Choi map.
\end{itemize}

In the next subsection, we will explore how a specific positive map, determined by the configurations of $\mu$ and $\Phi$ can be utilized to detect various quantum information theoretic resources, including Schmidt number, and other higher-dimensional entangled states such as PPT entanglement.

\subsection{Detection of Schmidt number}

To detect states whose Schmidt number is greater than $r$, we consider moments of $r$ positive but not $r+1$ positive Reduction map throughout this subsection. By considering different $n$-th order moments, we propose the following theorems for identifying states with Schmidt numbers exceeding $r$.\\
\\
\textbf{Theorem 1:} If a bipartite quantum state $\rho_{AB}$ has Schmidt number at most $r$, then the following inequality holds:
\begin{equation}
s_2^2 \leq s_3, \label{SN_criteria}
\end{equation}
where $s_2$ and $s_3$ are the second and third order moments corresponding to the $r$ positive but not $r+1$ positive Reduction map $\Lambda_R$ as defined in Eq.~\eqref{moments}.

\proof  Let, $\rho_{AB} \in  \mathcal{D}({{\mathbf{C}}^d \otimes  {\mathbf{C}}^d})$ be a bipartite quantum state which has Schmidt number at most $r$. Let $\Lambda_R$ be a $r$ positive but not $r+1$ positive Reduction map as defined in Eq. \eqref{reduction}. Using Eq.~\eqref{def_S}, we have 
\begin{equation}
     S_R= \frac{(id_A \otimes \Lambda_R)(\rho_{AB})}{\text{Tr}((id_A \otimes \Lambda_R)(\rho_{AB}))}.
\end{equation} 
From Ref.~\cite{terhal2000schmidt}, it is known that $(id_A \otimes {\Lambda_R} )(\rho_{AB}) \ge 0$ and hence, $S_R$ is a positive semidefinite matrix with unit trace. Let us now define Schatten-$p$ norms for $p \ge 1$ on the positive semidefinite operator $S_R$ as
 \begin{equation}
    ||S_R||_{p} := (\sum_{i=1}^{d}{|\chi_i|^p})^{\frac{1}{p}}=(\text{Tr}[|S_R|^p])^{\frac{1}{p}}   \label{schattenp}
\end{equation}
 where $S_R$ has the spectral decomposition $S_R= \sum_{i=1}^{d} \chi_{i} \ket{x_i}\bra{x_i}$. 
 Further, the $l_p$ norm of the vector of eigenvalues of $S_R$ corresponding to each Schatten-$p$ norm is defined by:
  \begin{equation}
     || \chi||_{l_p} := (\sum_{i=1}^{d}{|\chi_i|^p})^{\frac{1}{p}} \label{lpnorm}
 \end{equation}
 where ${\{\chi_i \}_{{i=1}}^{d}}$ is the set of eigenvalues of $S_R$.
 The inner product of two vectors in $\mathbb{R}^{d}$ is defined as
 \begin{equation}
     \langle u, v \rangle :=\sum_{i=1}^{d} u_{i} v_{i} \label{innerproduct}
 \end{equation}
 for $u,v \in \mathbb{R}^{d}$.
 Now, from H\"{o}lder's inequality for vector norms, we know that for $p, q \ge 1$ and $\frac{1}{p}+\frac{1}{q} =1,$
 the following relation holds:
 \begin{equation}
     |\langle u, v\rangle | \le \sum_{i=1}^{d} |u_{i} v_{i}| \le ||u||_{l_p} ||v||_{l_q} .\label{hoelderinequality}
 \end{equation}
 Putting $p=3$, $q=\frac{3}{2}$ and $u=v= \chi$ in \eqref{hoelderinequality}, we get
 \begin{equation}
     \text{Tr}[{S_R}^2]= \langle\chi, \chi\rangle \le ||\chi||_{l_3} ||\chi||_{l_{\frac{3}{2}}}=||S_R||_{3} ||\chi||_{l_{\frac{3}{2}}}.
     \label{applyinghoelder}
 \end{equation}
 Note that the Cauchy-Schwarz inequality is obtained by putting $p=\frac{1}{2}$ and $q=\frac{1}{2}$ in H\"{o}lder's inequality. \\
 Now, 
 \begin{align}
&{||S_R||_{2}}^2=\text{Tr}[{S_R}^2] \nonumber \\ 
& ~~~~~ \overset{a}{\leq} ||S_R||_{3} ||\chi||_{l_{\frac{3}{2}}} \nonumber \\ 
&   ~~~~~ =  ||S_R||_{3} (\sum_{i=1}^{d}{|\chi_i|^{\frac{3}{2}}})^{\frac{2}{3}} \nonumber \\ 
&     ~~~~~ =  ||S_R||_{3} (\sum_{i=1}^{d}{{|\chi_i|}{|\chi_i|}^{\frac{1}{2}}})^{\frac{2}{3}} \nonumber \\ 
&     ~~~~~  \overset{b}{\leq}  ||S_R||_{3}  ((\sum_{i=1}^{d}{|\chi_i|^2})^{\frac{1}{2}} (\sum_{i=1}^{d}{|\chi_i|})^{\frac{1}{2}})^{\frac{2}{3}}\nonumber \\ 
&     ~~~~~ =||S_R||_{3}  {||S_R||_{2}}^{\frac{2}{3}} {||S_R||_{1}}^{\frac{1}{3}}
 \label{applyingcauchy}
\end{align}
where (a) follows from Eq.~\eqref{applyinghoelder}, and (b) follows from the Cauchy–Schwarz inequality.
Taking $3$rd power of \eqref{applyingcauchy}, we get
\begin{equation}
   {||S_R||_{2}}^4 \le  {||S_R||_{3}}^3 ||S_R||_{1} . \label{applying3rdpower}
\end{equation}
Since $\text{Tr}(S_R) =1$, it follows that it follows that the trace norm of $S_R$ satisfies $||S_R||_{1} =1$, and hence reducing Eq.~\eqref{applying3rdpower} to:
\begin{equation}
   {||S_R||_{2}}^4 \le  {||S_R||_{3}}^3  \label{normproof}
\end{equation}
{\it i.e.},
\begin{equation}
   {s_2}^2 \leq s_3 \label{theorem1proof}
\end{equation}
which completes the proof. \qed \\

The above theorem implies that the condition presented in Eq.~\eqref{SN_criteria} is necessary for a state $\rho_{AB}$ to have Schmidt number at most $r$. Hence, violation of the above theorem is sufficient to conclude that the Schmidt number of the quantum state is greater than $r$. \\

We now exploit higher order moments to propose another criterion capable of efficiently detecting states with Schmidt number greater than $r$. \\
\\
\textbf{Theorem 2:} If a bipartite quantum state $\rho_{AB}$ has Schmidt number less than or equal to $r$, then

\begin{equation}
       \det[H_{m}(\mathbf{S}_R)] \ge 0 . \label{Hankelmatrix}
    \end{equation}
Here, $[H_{m}(\mathbf{s})]_{ij} = s_{i+j+1}$ for $i,j \in \{0,1,...,m\}$, $m \in \mathbb{N}$ and $s_i$, $i= 1,2,..,n$  are the i-th moments defined in Eq.~\eqref{moments} corresponding to $r$ positive but not $r+1$ positive Reduction map $\Lambda_R$.

\proof  Let, $\rho_{AB} \in  \mathcal{D}({{\mathbf{C}}^d \otimes  {\mathbf{C}}^d})$ be a bipartite quantum state which has Schmidt number less than or equal to $r$. We consider the $r$ positive but not $r+1$ positive Reduction map $\Lambda_R$ defined in Eq.~\eqref{reduction}. As mentioned earlier, $S_R$ is a positive semidefinite matrix with unit trace and therefore if  $S_R= \sum_{i=1}^{d} \chi_{i} \ket{x_i}\bra{x_i}$ be the spectral decomposition of $S_R$, then $\chi_{i} \ge 0$ for $i=1,2,..,d$.

Now, if $\mathbf{S}_R = (s_1,s_2,....s_n)$ be the moment vector defined in Eq. \eqref{moments}, then the $(m+1) \times (m+1)$ Hankel matrices are given by the elements $[H_{m}(\mathbf{S}_R)]_{ij} = s_{i+j+1},$ with $i,j \in \{ 0,1, ..., m\}$.

These Hankel matrices $H_{m}(\mathbf{S}_R)$ can also be written as 
  \begin{equation}
      H_{m}(\mathbf{S}_R) = V_m D V_m^T
  \end{equation}
where,\begin{equation}
    V_m = \begin{pmatrix}
1 & 1 & ...&1  \vspace{0.2cm}\\ 
\chi_1 & \chi_2 & ...& \chi_d  \vspace{0.2cm}\\
...&...&...&...&\\
...&...&...&...&\\
\chi_1^m & \chi_2^m & ...& \chi_d^m      \label{secondHankelmatrix}
\end{pmatrix}
\end{equation} 
and \begin{equation}
    D = \begin{pmatrix}
 \chi_1& 0 & ...&0  \vspace{0.2cm}\\ 
0 & \chi_2 & ...& 0  \vspace{0.2cm}\\
...&...&...&...&\\
...&...&...&...&\\
0 & 0 & ...&   \chi_d  \label{secondHankelmatrix}
\end{pmatrix}.
\end{equation} 

Now, for an arbitrary vector $x=(x_1,...x_m,x_{m+1}) \in {\mathbb{R}}^{m+1}$, we have 
\begin{equation}
    x  H_{m}(\mathbf{S}_R) x^T = x  V_m D V_m^T  x^T = y D y^T = \sum_{i=1}^d \chi_i {y_i}^2 \ge 0,
\end{equation}
where, $y= x  V_m = (y_1,y_2,...y_d) $ with $y_i= \sum_{j=1}^{m+1} x_j \hspace{0.1cm}{\chi_i}^{j-1}, \hspace{0.1cm} \text{ for } i=1,2,....d$. \\
Hence, $ x  H_{m}(\mathbf{S}_R) x^T \ge 0$ which implies $ H_{m}(\mathbf{S}_R) \ge 0$, {\it i.e.} $\det[H_{m}(\mathbf{S}_R)] \ge 0$. This completes the proof. \qed\\

Similar to Theorem 1, the above theorem confirms that the condition in Eq.~\eqref{Hankelmatrix} is a necessary requirement for a state to have a Schmidt number of at most $r$. Therefore, any violation of this condition is sufficient to conclude that the Schmidt number of the quantum state is greater than $r$.

\subsubsection{Examples:}
We now introduce several examples that illustrate and support our proposed detection criteria.\\
\\
\textbf{Example 1}: Consider a qutrit isotropic state of the form:
\begin{equation}
    \rho_{iso}= p \ket{{\phi}^{+}_3} \bra{{\phi}^{+}_3} + \frac{1-p}{9}  I_9
\end{equation}
where, $\ket{{\phi}^{+}_3}=\frac{1}{\sqrt 3} \sum_{i} \ket{ii}$ is the maximally entangled state in ${\mathbf{C}}^3\otimes {\mathbf{C}}^3$ and $p \in [0,1]$. As shown in Ref.~\cite{terhal2000schmidt}, the Schmidt number of this state is at most $2$ if and only if (iff) $0 \le p \le \frac{5}{8}$ and SN $(\rho_{iso}) =3$ iff $\frac{5}{8} < p \le 1$. Applying our criterion proposed in Theorem 1, we find that the condition ${s_2}^2 > s_3$ holds precisely in the range $\frac{5}{8} < p \le 1$ for a $2$ positive but not $3$ positive Reduction map. Hence, our criteria proposed in Theorem 1 can successfully detect the entire range of the parameter $p$ for which SN$( \rho_{iso}) =3$.\\
\\
\textbf{Example 2}: Consider the scenario where the dephasing map in the computational basis is applied to a maximally entangled state, resulting in a noisy state of the form:
\begin{equation}
    \rho_{dep}= v \ket{{\phi}^{+}_3} \bra{{\phi}^{+}_3} + \frac{1-v}{3}  \sum_{i=0}^2 \ket{ii} \bra{ii} 
\end{equation}
where, $\ket{{\phi}^{+}_3}$ is the maximally entangled state in ${\mathbf{C}}^3\otimes {\mathbf{C}}^3$ as defined earlier and $v \in [0,1]$. Ref.~\cite{terhal2000schmidt} shows that this state has Schmidt number at most $2$ iff $ 0 \le v \le \frac{1}{2}$ and SN $(\rho_{dep}) =3$ iff $\frac{1}{2} < v \le 1$.

However, if we now apply our criterion proposed in Theorem 1, we obtain ${s_2}^2 > s_3$ only for the parameter range $0.56 \le v \le 1$ for a $2$ positive but not $3$ positive Reduction map. Therefore, our criteria proposed in Theorem 1 can not detect the full range of the parameter $v$ for which SN$( \rho_{dep}) =3$.\\

Next, we apply our proposed criterion from Theorem 2 to detect Schmidt number of $\rho_{dep}$. It is important to note here that this criterion is violated in the exact parameter region in which Schmidt number of $\rho_{dep}$ is $3$ {\it i.e.} $\det[H_{2}(\mathbf{S_2})]$ is not positive semidefinite for a $2$ positive but not $3$ positive Reduction map in the parameter region $0.5 < v \le 1$. Hence, Theorem 2 provides a tighter condition to detect Schmidt number for $\rho_{dep}$.


\subsection{Detection of PPT and NPT entangled states} 
In the previous section, we utilized moments of $r$ positive and not $r+1$ positive Reduction map to certify quantum states with a Schmidt number greater than $r$.  In this subsection, we extend our approach by considering moments of other positive maps, including the Choi map, the Breuer–Hall map, and the Reduction map, to detect both PPT and NPT entangled states.\\

\textbf{Theorem 3:} If a bipartite quantum state is separable, then 

\begin{equation}
       \det[H_{m}(\mathbf{S})] \ge 0 . \label{Hankelmatrixcondition}
    \end{equation}
Here, $ [H_{m}(\mathbf{S})]_{ij} = s_{i+j+1}$, and $s_i$ with $i= \{1,2,..,n\}$ are defined in Eq.~\eqref{moments} corresponding to the Reduction, Choi and the Breuer–Hall map.

\proof Let, $\rho_{AB} \in  \mathcal{D}({{\mathbf{C}}^d \otimes  {\mathbf{C}}^d})$ be a separable quantum state. Here we consider the Choi, Reduction and the Breuer-Hall map defined in Eq.~\eqref{reduction}. Now, if we define \begin{equation}
     S= \frac{(id_A \otimes \Lambda_{\mathcal{x}})(\rho_{AB})}{\text{Tr}((id_A \otimes \Lambda_{\mathcal{x}})(\rho_{AB}))}.
\end{equation} 
Then from the properties of positive map \cite{horodecki1996necessary}, we obtain that $S$ is a positive semidefinite operator with unit trace. Therefore, if $\{\nu_i\}_{i=1}^d$ are the eigenvalues of $S$, then $\nu_i \ge 0$ for $i=1,2,..,d$.

Let us now denote $\mathbf{S} = (s_1,s_2,....s_n)$ be the moment vector defined in Eq. \eqref{moments}. We can define the $(m+1) \times (m+1)$ Hankel matrices with elements $[H_{m}(\mathbf{S})]_{ij} = s_{i+j+1},$ for $i,j \in \{ 0,1, ..., m\}$. 
The remainder of the proof follows similarly to Theorem 2 by replacing $\mathbf{S}_R$ by $\mathbf{S}$. \qed

\subsubsection{Examples of PPT entangled states:} We now present examples of PPT entangled states which can be detected by the moments of the Choi and the Breuer-Hall map.\\
\\
\textbf{Example 3}: Consider a positive partial transpose (PPT) entangled state in $\mathbf{C}^3 \otimes \mathbf{C}^3$ defined as follows \cite{stormer1982decomposable,bhattacharya2021generating},
\begin{equation}
    \rho_{\text{bound}} = \frac{1}{1+p+\frac{1}{p}}\begin{bmatrix} 
	1& 0 &0 & 0& 1& 0&0&0&	1\\[0.1cm]
 0& p &0 & 0& 0& 0&0&0&	0  \\[0.1cm]
	0& 0 &\frac{1}{p} & 0& 0& 0&0&0&	0\\[0.1cm]
	0& 0 &0 & \frac{1}{p}& 0& 0&0&0&	0\\[0.1cm]
    1& 0 &0 & 0& 1& 0&0&0&	1\\[0.1cm]
    0& 0 &0 & 0& 0& p&0&0&	0\\[0.1cm]
    0& 0 &0 & 0& 0& 0&p&0&	0\\[0.1cm]
    0& 0 &0 & 0& 0& 0&0&\frac{1}{p}&	0\\[0.1cm]
    1& 0 &0 & 0& 1& 0&0&0&	1\\[0.1cm]
	\end{bmatrix}\\
\end{equation}
where, $p$ is a non-zero, positive, real number. Here, we use the moments of the Choi map, which was previously introduced in \cite{wang2022operational}. Below, we illustrate that these moments can be used to detect the PPT entanglement of the state mentioned above.

By applying the criterion proposed in Theorem 3, we find that $ \det[H_{2}(\mathbf{S})] <0$ for the parameter range $p \in [0.06,1)$. This result demonstrates that the moments of the Choi map can successfully detect the bound entangled state  $\rho_{\text{bound}}$.\\

\textbf{Example 4}: Bennett {\it et al.}~\cite{bennett1999unextendible} introduced the concept of an unextendable product basis (UPB). The set $\mathcal{U}_{\textit{tiles}} = \{\ket{u_1},\ket{u_2},\ket{u_3},\ket{u_4},\ket{u_5}\} \subset \mathbf{C}^3 \otimes \mathbf{C}^3$, is commonly referred as the \textit{tiles} UPB, where 
\begin{equation}
\begin{split}
   & \ket{u_1} = \ket{0}\otimes\frac{\ket{0}- \ket{1}}{\sqrt{2}},   \ket{u_2} = \ket{2}\otimes \frac{\ket{1}- \ket{2}}{\sqrt{2}} \\
 & \ket{u_3} = \frac{\ket{0}- \ket{1}}{\sqrt{2}} \otimes \ket{2},   \ket{u_4} = \frac{\ket{1}- \ket{2}}{\sqrt{2}} \otimes \ket{0},\\
 & \ket{u_5} = \frac{1}{3} (\ket{0}+ \ket{1} +\ket{3}) \otimes (\ket{0}+ \ket{1} +\ket{3}).
   \end{split}
    \end{equation}
 Since no product state lies in the orthogonal complement of these states~\cite{bennett1999unextendible},  therefore, the state 
\begin{equation}
   \rho_{\textit{tiles}} = \frac{1}{4} (\mathbf{I}_9- \sum_{i=1}^5 \ket{u_i} \bra{u_i})
\end{equation}
is entangled. Moreover, due to its construction, this state has a positive partial transpose (PPT), making it an example of a PPT entangled state.

Applying the criterion proposed in Theorem 3 to the moments of the Choi map yields $\det[H_{2}(\mathbf{S})] =4.65023 \cross 10^{-8}\ge 0$, indicating that these moments, as defined in Eq.~\eqref{moments}, fail to detect $\rho_{\text{tiles}}$. However, when the same criterion is applied to the moments of the Breuer–Hall map, we obtain $\det[H_{2}(\mathbf{S})] = -0.00112943 <0$,  using the antisymmetric matrix \begin{equation}
   U = \begin{bmatrix} 
	 0 &-1 & 0\\[0.1cm]
 1& 0 &0  \\[0.1cm]
	0& 0 &0 \\[0.1cm]
	\end{bmatrix}\\
	\end{equation} Hence the criteria proposed in Theorem 3 corresponding to moments of the Breuer–Hall map, is capable of detecting $\rho_{\textit{tiles}}$.
\\
\subsubsection{Examples of NPT entangled states:}
\textbf{Example 5}: Consider a class of NPT entangled states in $\mathbf{C}^3 \otimes \mathbf{C}^3$ defined as follows \cite{garg2021teleportation}:
\begin{equation}
    \rho_{\text{NPT}} = \begin{bmatrix} 
	\frac{1-\alpha}{2}& 0 &0 & 0& 0& 0&0&0&	-\frac{11}{50}\\[0.1cm]
 0& 0 &0 & 0& 0& 0&0&0&	0  \\[0.1cm]
	0& 0 &0 & 0& 0& 0&0&0&	0\\[0.1cm]
	0& 0 &0 & 0& 0& 0&0&0&	0\\[0.1cm]
    0& 0 &0 & 0& \frac{1}{2}-\alpha& -\frac{11}{50}&0&0&	0\\[0.1cm]
    0& 0 &0 & 0& -\frac{11}{50}& \alpha&0&0&	0\\[0.1cm]
    0& 0 &0 & 0& 0& 0&0&0&	0\\[0.1cm]
    0& 0 &0 & 0& 0& 0&0&0&	0\\[0.1cm]
    -\frac{11}{50}& 0 &0 & 0& 0& 0&0&0&	\frac{\alpha}{2}\\[0.1cm]
	\end{bmatrix}\\
	\end{equation}
where, $\frac{25-\sqrt{141}}{50} \le \alpha \le \frac{25+\sqrt{141}}{100}$.

Although the state mentioned above was detected using partial realigned moments in \cite{aggarwal2024entanglement}, we demonstrate here that it can also be detected using the moments of the Reduction map. Applying the criterion from Theorem 3 to the moments of the Reduction map with $k=1$ yields $\det[H_{2}(\mathbf{\mathbf{S}})] < 0$ for the full range of $\alpha$. Thus, our criterion, based on the moments of the Reduction map defined in Eq.~\eqref{moments}, successfully detects $\rho_{\text{NPT}}$.

\section{Detection of quantum channels and their discrimination}\label{s4} 
In this section, we investigate the implications of our proposed criteria in quantum communication channels using the moments of generalized positive maps.

In the context of quantum communication, there exists a class of resource-breaking channels referred to as Schmidt number breaking channels, that reduce the Schmidt number of bipartite states, thereby diminishing the effective dimensionality of entanglement. Detecting channels that are not Schmidt number breaking is therefore essential for preserving higher-dimensional entanglement, an important resource for quantum communication, distributed computing, and related applications. Motivated by this, in the following subsection we focus on the detection of non–Schmidt number breaking channels, using our moment-based criteria. 
We then highlight additional operational implications of our approach in quantum channel discrimination tasks.

\subsection{Detection of non-Schmidt number breaking channels}
In this subsection, we also utilize the moments of the $k$-Reduction map, defined in Eq.~\eqref{moments} to identify channels which are not Schmidt number-breaking.\\
\\
\textbf{Definition 2:} Let $\mathcal{E}: \mathcal{M}_d \rightarrow \mathcal{M}_d$ be a quantum channel. The moments of the Schmidt number breaking channels ($\mathbb{SNBC}$) are defined as 
\begin{equation}
e_n = \text{Tr}[{E_R}^n] \label{channelmoments}
\end{equation}
where, \begin{equation} \label{E_R}
  E_R = \frac{(id_A \otimes \Lambda_R)(id_A \otimes {\mathcal{E}}) \ket{{\phi}^+} \bra{{\phi}^+})}{\text{Tr}((id_A \otimes \Lambda_R)(id_A \otimes {\mathcal{E}}) \ket{{\phi}^+} \bra{{\phi}^+}))}
\end{equation}  
with $n$ being a positive integer and $\Lambda_R$ is a $r$-positive but not $r+1$ positive Reduction map defined in Eq.~\eqref{reduction}.

Utilizing only the second and third order moments of the channels {\it i.e.}, $e_2$ and $e_3$, we propose our first criterion, presented as a theorem below.\\
\\
\textbf{Theorem 4:} If a quantum channel $\mathcal{E}: \mathcal{M}_d \rightarrow \mathcal{M}_d$ has Schmidt number at most $r$, then the following inequality holds:
\begin{equation}
e_2^2 \leq e_3, \label{6}
\end{equation}
where $e_2$ and $e_3$ are defined in Eq.~\eqref{channelmoments} corresponding to $r$ positive Reduction map.\\
\proof Let, $\mathcal{E}: \mathcal{M}_d \rightarrow \mathcal{M}_d$ be a quantum channel which has Schmidt number less than or equal to $r$ {\it i.e.} SN$[{\mathcal{C}}_{\mathcal{E}}] \le r$, with ${\mathcal{C}}_{\mathcal{E}}$ being the Choi operator corresponding to the channel ${\mathcal{E}}$. If $E_R$ is obtained from Eq.~\eqref{E_R} corresponding to a $r$ positive but not $r+1$ positive Reduction map $\Lambda_R$, then from Ref. \cite{terhal2000schmidt}, we know that $E_R$ is a positive semidefinite operator with unit trace. The remainder of the proof then follows analogously to Theorem 1, replacing $S_R$ by $E_R$. \qed\\

We now utilize higher-order moments to propose a refined criterion for efficiently detecting quantum channels which have Schmidt number greater than $r$.\\

\textbf{Theorem 5:} If a quantum channel $\mathcal{E}: \mathcal{M}_d \rightarrow \mathcal{M}_d$ has Schmidt number at most $r$, then
\begin{equation}
      \det[H_{n}(\mathbf{E})] \ge 0 . \label{Hankelmatrixcondition}
    \end{equation}
Here, $[H_{n}(\mathbf{E})]_{ij} = e_{i+j+1}$, are the Hankel matrices and $e_i$ with $i= 1,2,..,n$  are the channel moments corresponding to the $r$ positive Reduction map defined in Eq.~\eqref{channelmoments}.
\proof Consider a quantum channel $\mathcal{E}: \mathcal{M}_d \rightarrow \mathcal{M}_d$ which has Schmidt number atmost $r$ {\it i.e.} Schmidt number of its corresponding Choi operator ${\mathcal{C}}_{\mathcal{E}}$ satisfies SN$[{\mathcal{C}}_{\mathcal{E}}] \le r$. 
Note that $E_R$ as defined in Eq.~\eqref{E_R} corresponding to the $r$ positive but not $r+1$ positive Reduction map $\Lambda_R$ (defined in Eq. \eqref{reduction}), is a positive semidefinite operator with unit trace having spectral decomposition $E_R= \sum_{i=1}^{d} \lambda_{i} \ket{x_i}\bra{x_i}$ with $\lambda_{i} \ge 0$ for $i=1,2,..,d$.

If $\mathbf{E}_R = (e_1,e_2,....e_n)$ is the moment vector defined in Eq. \eqref{channelmoments}, then we can construct the $(m+1) \times (m+1)$ Hankel matrices with the elements $[H_{m}(\mathbf{E}_R)]_{ij} = e_{i+j+1},$ (where $i,j \in \{ 0,1, ..., m\}$). Rest of the proof proceeds similarly to Theorem 2 with $\mathbf{S}_R$ replaced by $\mathbf{E}_R$. \qed\\

Analogous to Theorem 4, any violation of this condition implies that the Schmidt number of the quantum channel is greater than $r$.

\subsubsection{Examples:} 
We now present two examples in support of our detection schemes.\\
\\
\textbf{Example 6} (\textit{Depolarizing channel}): Let us first consider the depolarizing channel $  {\mathcal{K}}_d : \mathcal{M}_d \rightarrow \mathcal{M}_d $ whose action is given by   \cite{nielsen2010quantum,chruscinski2006partially},
\begin{equation} \label{depolarzing}
    {\mathcal{K}}_d(\rho) = p \rho + \frac{1-p}{d} \text{Tr}(\rho) I_d
\end{equation}
where, $p\in [0,1]$. It is known from Ref.~\cite{terhal2000schmidt} that SN$( {\mathcal{K}}_d) \le r$ iff 
\begin{equation} \label{SNphi}
  0 \le  p \le  \frac{rd-1}{d^2-1}.
\end{equation}

For simplicity, let us take $d=3$. Then Eq. \eqref{depolarzing} becomes
\begin{equation}
   {\mathcal{K}}_3 (\rho) = p \rho + \frac{1-p}{3} \text{Tr}(\rho) I_3
\end{equation}

Note that, ${\mathcal{K}}_3 \in$ $1-\mathbb{SNBC}$ i.e. $\mathbb{EB}$ iff $ 0 \le p \le  \frac{1}{4}$ and SN $({\mathcal{K}}_3) >1$ iff $  \frac{1}{4} < p \le 1 $. Using Theorem 4, we obtain 
\begin{equation} \nonumber
   {e_2}^2 \leq e_3   \hspace{0.2cm} \text{for} \hspace{0.2cm} 0 \le p \le  \frac{1}{4}
\end{equation}
and 
\begin{equation} \nonumber
   {e_2}^2 > e_3 \hspace{0.2cm} \text{for} \hspace{0.2cm} \frac{1}{4} < p \le 1.
\end{equation} 
Hence, our proposed criterion in Theorem 4 can successfully detect the entire range where the channel ${\mathcal{K}}_3$ is not $\mathbb{EB}$.\\

Now from Eq. \eqref{SNphi}, it follows that $ {\mathcal{K}}_3 \in$ $2-\mathbb{SNBC}$ iff $ 0 \le p \le  \frac{5}{8}$ and SN $({\mathcal{K}}_3) =3$ iff $  \frac{5}{8} < p \le 1 $.
Using our proposed criteria defined in Theorem 4, we get 
\begin{equation}
   {e_2}^2 \leq e_3   \hspace{0.2cm} \text{for} \hspace{0.2cm} 0 \le p \le  \frac{5}{8}
\end{equation}
and 
\begin{equation}
   {e_2}^2 > e_3 \hspace{0.2cm} \text{for} \hspace{0.2cm} \frac{5}{8} < p \le 1.
\end{equation}
Hence, our criteria proposed in Theorem 4 can detect the entire range of the parameter $p$ for which ${\mathcal{K}}_3 \notin 2-\mathbb{SNBC}$. One may check that indeed for the parameter regime $\frac{5}{8} < p \le 1$, SN$({\mathcal{K}}_3) =3$.\\
\\
\textbf{Example 7} (\textit{Dephasing channel}): We now consider the dephasing channel $  {\mathcal{P}}_d : \mathcal{M}_d \rightarrow \mathcal{M}_d $, defined by its action as follows \cite{nielsen2010quantum,tavakoli2024enhanced}:
\begin{equation} \label{dephasing}
    {\mathcal{P}}_d(\rho) = v \rho + \frac{1-v}{d} \sum_{i=0}^{d-1} \ket{i}\bra{i}
\end{equation}
where, $v\in [0,1]$. Ref. \cite{tavakoli2024enhanced,bavaresco2018measurements} established that the Schmidt number of the dephasing channel, SN$( {\mathcal{P}}_d) \le r$ iff
\begin{equation} \label{SND}
   0 \le v \le \frac{r-1}{d-1}.
\end{equation}

We focus on the qutrit dephasing channel by setting $d=3$ in Eq. \eqref{dephasing}.\\

One may Note that ${\mathcal{P}}_d$ is $\mathbb{EB}$ iff $ v=0$ and not $\mathbb{EB}$ iff $ (0,1]$. On the other hand, Theorem 4 gives us \begin{equation} \nonumber
   {e_2}^2 \leq e_3   \hspace{0.2cm} \text{for} \hspace{0.2cm} v=0
\end{equation}
and 
\begin{equation} \nonumber
   {e_2}^2 > e_3 \hspace{0.2cm} \text{for} \hspace{0.2cm} v \in (0,1].
\end{equation} 
Therefore, the criterion proposed in Theorem 4 can accurately identify the entire range in which the channel 
 ${\mathcal{P}}_3$ is not $\mathbb{EB}$.\\

From Eq. \eqref{SND}, it follows that $ {\mathcal{P}}_3 \in 2-\mathbb{SNBC}$ iff $ 0 \le v \le  \frac{1}{2}$ and SN $( {\mathcal{P}}_3) =3$, {\it i.e.} $ {\mathcal{P}}_3 \notin 2-\mathbb{SNBC}$ iff $  \frac{1}{2} < v \le 1 $. \\

Now, applying our criterion from Theorem 4 yields the condition ${e_2}^2 > e_3$ for the parameter range $0.56 \le v \le 1$ when using the $2$ positive but not $3$ positive Reduction map. Thus, Theorem 4 does not capture the full range of $v$ for which SN$( \rho_{dep}) =3$.

We then apply our criterion from Theorem 5 to identify the parameter range where ${\mathcal{P}}_3$ is non-Schmidt number breaking channel. Notably, the criterion defined in Theorem 5 is violated exactly in the region where the Schmidt number of $ {\mathcal{P}}_3$ is 3 {\it i.e.} $\det[H_{2}(\mathbf{E})]$ fails to be positive semidefinite for $0.5 < v \le 1$. This demonstrates that Theorem 5 provides a tighter condition to detect Schmidt number of $ {\mathcal{P}}_3$.

\subsection{Operational implication of moment criteria in channel discrimination tasks} 
 In this subsection, we examine the operational implications of our proposed moment criteria in the context of channel discrimination tasks.

The task of channel discrimination is closely related to the well-studied problem of quantum state discrimination \cite{piani2009all}, which has been briefly discussed in Sec.~\ref{s2D}. Depending on the specific requirements of a given quantum information processing task, one channel may confer a greater operational advantage than the other. Therefore, the ability to distinguish between two quantum channels becomes a task of both theoretical and practical significance \cite{christandl2009postselection,macchiavello2013quantum}.  Consider the simplest scenario of discriminating two quantum channels $\mathcal{S}_i$ : $\mathcal{M}_d \rightarrow  \mathcal{M}_d$, for $i \in \{1,2\}$, each occurring with probability $p$ and $1-p$ respectively. The goal is to correctly identify which channel (\textit{i.e.}, the value of $i$) is applied while minimizing the probability of error. 

For a fixed input state $\rho$, the corresponding output states are $\mathcal{S}_1 (\rho)$ and $\mathcal{S}_2 (\rho)$. Now, the problem effectively reduces to the task of discriminating between the two states $\mathcal{S}_1 (\rho)$ and $\mathcal{S}_2 (\rho)$.  However, we can improve the discrimination process by introducing an ancillary system, as probe-ancilla entanglement can increase the success probability \cite{kitaev1997quantum,d2001using}. By preparing a state $\rho_{AR} \in \mathcal{D}(\mathbf{C}^d \otimes \mathbf{C}^r)$ with a $r$-dimensional ancilla such that the channel acts on the system alone, the output states become $\rho_i= (id_r \otimes \mathcal{S}_i) (\rho_{AR})$, $i \in \{1,2\}$. 
Unlike Eq.~\eqref{minimumerror} which requires optimization over all possible generalized measurements, this scenario involves the optimization over all input states in $\mathcal{D}(\mathbf{C}^d \otimes \mathbf{C}^r)$, to ensure that the resulting output states $\rho_i$ are distinguishable with minimum error for the given channels. This is mathematically captured by introducing a family of norms on Hermitian maps defined by
\begin{equation}
    ||\mathcal{S}||^r = \max_{\rho_{AR}}  || id_r \otimes \mathcal{S} (\rho_{AR})||_1.
\end{equation}

The distance between the two channels $\mathcal{S}_1$ and $\mathcal{S}_2$ (occurring with probability $p$ and $1-p$) when optimized over all input quantum states can be written as
\begin{equation}
    \mathbb{D}^r (\{\tilde{\mathcal{S}_1},\tilde{\mathcal{S}_2} \}) = || \tilde{\mathcal{S}_1}-\tilde{\mathcal{S}_2}||^r
\end{equation}
for $\tilde{\mathcal{S}_1} = p {\mathcal{S}_1}$ and $\tilde{\mathcal{S}_2} = (1-p) {\mathcal{S}_2}$. The optimal guessing probability (optimized over the choice of input states and that of final measurements) for the two quantum channels using an $r$-dimensional ancillary system is then given by
\begin{equation}
      p_{success} (\{ \tilde{\mathcal{S}_1}, \tilde{\mathcal{S}_2}\}) = \frac{1}{2} (1+ || \mathbb{D}^r (\{ \tilde{\mathcal{S}_1},\tilde{\mathcal{S}_2} \}) ||_1 ).
\end{equation}

With this, our aim is now to present the operational implications of our proposed moment criteria in channel discrimination task, which we provide below as a theorem.\\

\textbf{Theorem 6:} If for a bipartite quantum state $\rho_{AR}$, 
\begin{equation}
       \det[H_{m}(\mathbf{S}_R)] < 0  \label{Hankelmatrixcondition}
    \end{equation}
where, $[H_{m}(\mathbf{s})]_{ij} = s_{i+j+1}$ for $i,j \in \{0,1,...,m\}$, $m \in \mathbb{N}$ and $s_i$, $i= 1,2,..,n$  are the i-th moments defined in Eq.~\eqref{moments} corresponding to $r$ positive but not $r+1$ positive Reduction map $\Lambda_R$, then there exists two quantum channels $\mathcal{S}_1$, $\mathcal{S}_2 : \mathcal{M}_d \rightarrow  \mathcal{M}_d$ such that 
\begin{equation}
  \frac{1}{2}  || id_A \otimes \mathcal{S}_1 (\rho_{AR}) - id_A \otimes \mathcal{S}_2 (\rho_{AR}) ||_1 >  \mathbb{D}^r (\{ \frac{1}{2}\tilde{\mathcal{S}_1}, \frac{1}{2}\tilde{\mathcal{S}_2} \}).
\end{equation}

\proof 
If for a bipartite quantum state $\rho_{AR}$, $ \det[H_{m}(\mathbf{S}_R)] < 0$ where, $[H_{m}(\mathbf{s})]_{ij} = s_{i+j+1}$ for $i,j \in \{0,1,...,m\}$, $m \in \mathbb{N}$ and $s_i$, $i= 1,2,..,n$  are the i-th moments defined in Eq.~\eqref{moments} corresponding to $r$ positive but not $r+1$ positive Reduction map $\Lambda_R$, then from Theorem 2, we can conclude that the state  $\rho_{AR}$ has Schmidt number greater than $r$. 

Note that the $r$ positive but not $r+1$ positive reduction map ${\Lambda}_R$ defined in \eqref{reduction} is trace-preserving only for $d=2$. For dimensions $d>2$, we can modify this map to ensure trace preservation by employing a similar approach used in \cite{piani2009all}. Let ${\Lambda}_R ^{TP}:\mathcal{M}_d \rightarrow \mathcal{M}_d$ denote the modified trace-preserving $r$ positive but not $r+1$ positive reduction map. Now, from this trace-preserving map $({\Lambda}_R ^{TP})$, one can construct a trace-annihilating map $({\Lambda}_R ^{TA})$ as follows:
\begin{equation} \label{tpform}
    {\Lambda}_R ^{TA} (\rho) = {\Lambda}_R ^{TP} (\rho) - \text{Tr}(\rho) \ket{f}\bra{f}
\end{equation}
where, $\ket{f}\bra{f}$ is orthogonal to all elements of $\mathcal{M}_d$. From \cite{piani2009all}, we know that a trace-annihilating map $ {\Lambda}_R ^{TA}$, can always be expressed as a scaled difference between two quantum channels $ \mathcal{S}_1$ and $ \mathcal{S}_2$, i.e. 
\begin{equation}\label{jordan}
    \mathcal{S}_1 - \mathcal{S}_2 = k {\Lambda}_R ^{TA}
\end{equation}
where $k$ is a positive constant. Now, using Eq. \eqref{tpform} and Eq. \eqref{jordan}, we obtain
\begin{equation}
     \mathcal{S}_1 - \mathcal{S}_2 = k ({\Lambda}_R ^{TP}  - \text{Tr}(.) \ket{f}\bra{f})
\end{equation}

For any state $\sigma_{AR}$, which has Schmidt number less than or equal to $r$, we have 
\begin{align}
    & || id_A \otimes \mathcal{S}_1 (\sigma_{AR}) - id_A \otimes \mathcal{S}_2 (\sigma_{AR}) ||_1 \\ \nonumber
     =& || id_A \otimes (\mathcal{S}_1 - \mathcal{S}_2) (\sigma_{AR}) ||_1 \\ \nonumber
     =& k|| (id_A \otimes   {\Lambda}_R ^{TP})(\sigma_{AR})  - \text{Tr}_R(\sigma_{AR}) \otimes \ket{f}\bra{f})  ||_1\\ \nonumber
     =&  k(|| (id_A \otimes   {\Lambda}_R ^{TP})(\sigma_{AR})||_1 + 1)\\ \nonumber
     =&2k. \nonumber
\end{align}
 Additionally, if for a bipartite quantum state $\rho_{AR}$, we get $\det[H_{m}(\mathbf{S}_R)] < 0 $, we obtain
    \begin{align}
    & || id_A \otimes \mathcal{S}_1 (\rho_{AR}) - id_A \otimes \mathcal{S}_2 (\rho_{AR}) ||_1 \\ \nonumber
       =& k|| (id_A \otimes   {\Lambda}_R ^{TP})(\rho_{AR})  - \text{Tr}_R(\rho_{AR}) \otimes \ket{f}\bra{f})  ||_1\\ \nonumber
     =&  k(|| (id_A \otimes   {\Lambda}_R ^{TP})(\rho_{AR})||_1 + 1)\\ \nonumber
     >&2k. \nonumber
\end{align}
Since, for a bipartite quantum state $\rho_{AR}$, if $\det[H_{m}(\mathbf{S}_R)] < 0 $, then $|| (id_A \otimes   {\Lambda}_R ^{TP})(\rho_{AR})||_1 >1$ which proves the theorem. \qed

Therefore, the above theorem implies that if an unknown quantum state $\rho_{AR}$ violates Eq.~\eqref{Hankelmatrixcondition}, then $\rho_{AR}$ necessarily offers an advantage in discriminating some pair of quantum channels over all states with Schmidt number less than or equal to $r$. This result highlights the operational implication of our proposed criterion presented in Theorem 1.

\section{Conclusions}\label{s5}
Detecting higher-dimensional entanglement states are of foremost importance, as such states can significantly enhance the performance and capabilities of various quantum information processing tasks \cite{bae2019more,zhang2025quantum,kues2017chip,cerf2002security,Bhattacharya21}. In this work, we identify the signature of such higher dimensional entanglement using the moments of generalized positive map, which can be efficiently implemented in an experimental setup. While higher-dimensional entanglement can be characterized by various quantities, here we primarily focus on the Schmidt number of quantum states and specific classes of PPT and NPT entangled states in qutrit-qutrit systems. Note that, the Schmidt number is not merely a mathematically motivated concept, but holds a clear operational significance in entanglement theory, playing a pivotal role in the design of quantum information protocols \cite{bae2019more,huber2013weak,zhang2025quantum,kues2017chip}.
This underscores the need for efficient characterization methods, which is one of the central goals of our work.

We present several examples in support of our moment-based detection scheme. Next, we demonstrate how our moment-based criterion could be applied for detection of useful quantum channels such as non-Schmidt number breaking channels, illustrating our approach through examples of the depolarizing and dephasing channels. Lastly, as a direct application of our proposed moment-based criteria, we demonstrate their operational significance in quantum channel discrimination tasks.

Our proposed criteria rely on computing simple functionals without necessitating the evaluation of the full spectrum of the quantum state, and that can be efficiently implemented in real experiments using shadow tomography \cite{aaronson2018shadow,aaronson2019gentle,huang2020predicting,elben2020mixed}. Moreover, our moment-based approach offers a significant advantage in terms of scalability. While conventional full state tomography demands a number of measurements that scales exponentially with the size of the quantum system, our method requires only a polynomial number of copies to detect and characterize higher-dimensional entanglement.

Our study paves the way for several promising future research directions. An immediate open avenue is to investigate the classes of entangled states for which moment-based conditions can serve as both necessary and sufficient criteria. Moreover, our proposed moments of the generalized Choi map can be utilized to detect other higher-dimensional entangled states, presenting a promising direction for future research. Given the experimental feasibility of our proposed protocol, another important next step is to realize these detection schemes in practical experimental setups.
\\{\it Note added:} Recently, we became aware of a related independent work \cite{yi2025certifying} with a complementary emphasis. While the focus of Ref.~\cite{yi2025certifying} seems to be more on refining the entanglement dimensionality detection criteria based on reduction moments and their numerical simulations, the focus of our present work is inclined towards a slightly broader approach of detecting higher dimensional quantum states and channels based on generalized moments of positive maps and its operational utility in channel discrimination tasks.
\section{Acknowledgements}
B.M. and A.G.M. acknowledge Saheli Mukherjee for insightful discussions. B.M. also acknowledges the DST INSPIRE fellowship program for financial support. A.G.M. acknowledges the financial
support through the National Quantum Mission (NQM) of the
Department of Science and Technology, Government of India. N.G. acknowledges support from the DST-ANRF (SERB) MATRICS grant vide file number MTR/2022/000101.

\bibliography{main}

\end{document}